\documentclass[letterpaper]{jpconf}
\usepackage{graphicx}

\def\he4{$^4$He}
\def\Am3{\AA$^{-3}$}
\def\beq{\begin{equation}}
\def\eeq{\end{equation}}
\def\hee3{$^3$He}

\begin{document}
\title{The role of \hee3 impurities in the stress induced  roughening of superclimbing dislocations in solid \he4} 

\author{D. Aleinikava and A.B. Kuklov}
\address{Department of Engineering Science and Physics,
CUNY, Staten Island, NY 10314, USA}
\ead{Anatoly.Kuklov@csi.cuny.edu}

\begin{abstract}
We analyze the stress induced and thermally assisted roughening of a forest of superclimbing  dislocations in a Peierls potential in the presence of \hee3 impurities and randomly frozen in static stresses. It is shown that the temperature of the dip $T_d$ in the flow rate observed by Ray and Hallock (Phys.Rev. Lett. {\bf 105}, 145301 (2010)) is determined by the energy of \hee3 activation from dislocation core. However, it is suppressed by, essentially, the logarithm of the impurity fraction. The width of the dip is determined by inhomogeneous fluctuations of the stresses and is shown to be much smaller than $T_d$. 
\end{abstract}

\section{Introduction}
Strong interest in the supersolid state of matter in free space \cite{SFS} has been revived by the recent discovery of the torsional oscillator (TO) anomaly in solid \he4 \cite{KC}. While finding no supersolidity in the ideal \he4 crystal, {\it ab initio} quantum Monte Carlo simulations did find that some grain boundaries \cite{GB}, dislocations \cite{screw,sclimb} or crystal boundaries \cite{Ceperley} support low-d superfluidity spatially modulated by the surrounding lattice. In principle, a percolating network of superfluid dislocations \cite{Shevchenko} could explain the TO anomaly if the dislocation density is 3-4 orders of magnitude higher than it is expected to be in a slowly grown and well annealed crystal.
Consistent with such expectation is also a very small rate of the critical superflow through solid \he4 (occuring presumably along dislocations with superfluid cores) observed in the UMASS-Sandwich experiments \cite{Ray,Hallock_2010}. Thus the nature of the TO anomaly in solid \he4 remains unclear.

In the present work we focus on the very unexpected feature of the UMASS-Sandwich experiment \cite{Hallock_2010} -- the strong suppression of the supercritical flow rate $V_{cr}$ and then its  recovery in a  narrow range of temperatures. Such a feature occurs well below (about 10 times) the flow onset temperature $T_O \approx 0.5-0.6$K \cite{Hallock_2010}. In our recent work \cite{us_PRL_2011} it has been proposed that this feature can be due to the stress induced roughening of 
superclimbing dislocations supporting the superflow. The key question, though, is why the temperature for the anomaly is pretty much independent of dislocation
density and why the dip is so narrow despite disordered nature of the solid \he4. Here we are proposing the explanation within the  scenario where \hee3 impurities as well as internal stresses biasing superclimbing dislocations turn out to be crucial.
We will show that the dip-anomaly temperature $T_{d}$ is determined by the \hee3 activation energy $E_a \sim 0.5-0.8$K which is reduced by a large logarithm of \hee3 fraction.

\section{\hee3 precipitation on dislocation core}
Atoms of \hee3 provide pinning centers for dislocations. Thus, Frank's forest of superclimbing dislocations can be viewed as consisting of free segments of quantum superclimbing strings \cite{sclimb,us_PRL_2011}
of length $L$ equal to the average distance between \hee3 impurities absorbed on dislocations, provided $L$ is less than a typical distance between the network cross-links. Here we derive the equation for $L$ as a function
of temperature within the thermodynamical model including the total number of \hee3 atoms in the crystal bulk $N_b$ and the total number of the impurities $N_d$ absorbed on dislocations. It is natural to presume that cores of the dislocations constitute a small fraction $x_d <<1$ of the total sample. Free energy of such system can be written in the following form
\beq
G =E_a N_b +T N_b \ln (N_b/eN_{0b}) + T N_d \ln(N_d/eN_{0d}) , \quad N_b +N_d = N_3, 
\label{F}
\eeq
where $N_3$ is the total number of \hee3 atoms (with their fraction $X_3 = N_3 / N_{0b} <<1$); $N_{0b}$ denotes the total number of sites available for \hee3 atoms
in the bulk; $N_{0d}= x_d N_{0b}$ stands for the total number of sites available for \hee3 atoms on dislocations. In this model we presume that $N_{0d} > N_3$ so that
there is no need to consider the effect of the entropy reduction on dislocations when all \hee3 condense on dislocations. Including this effect, while introducing additional
technical difficulties, does not alter
the main conclusion of this work.

Varying $G$ with respect to $N_d$ as $ dG/dN_d =0$, while keeping $N_3$ fixed, and introducing the fractional \hee3 concentration on dislocations $X=L^{-1}= N_d/N_{0d}$, we find
\beq
L^{-1}=\frac{X_3}{x_d + \exp(-E_a/T)},
\label{L}
\eeq
where the inter-impurity distance $L$ along a dislocation core is measured in units of a typical inter-atomic distance $b\approx 3-4$\AA  ~in solid \he4.  In what follows
we will be using this value $L=L(T)$ as a length of dislocation segments undergoing the stress induced roughening \cite{us_PRL_2011}.
  
\section{Stress induced roughening of a dislocation}
Superclimbing dislocation is modeled as a quantum string oriented along the $x$-axis and strongly pinned at its both ends $x=0,L$ \cite{Granato}. Here we will be considering
$L$ determined by the mean distance between \hee3 impurities absorbed on dislocations, Eq.(\ref{L}).
The string displacement $y(x,t)$ along  the $y$-axis depends on the time $t$ and is measured in units of the  inter-atomic spacing ($\approx$ Burger's vector $b$) with respect to its equilibrium $y=0$ (no tilting is considered). The Peierls potential induced by the crystal is taken as $U_P=-u_P \cos\left(2\pi y(x,t)\right)$. 
The partition function $Z$ has the form \cite{sclimb,JLTP,us_PRL_2011}   
\begin{eqnarray} 
Z&=&\int Dy(x,t)\, D\rho(x,t) D\phi(x,t) \exp(-S), 
\label{Z} \\
S&=& \int_0^\beta dt\sum_{x}[ i(\rho +n_0)\nabla_t \phi  + \frac{\rho_0}{2} (\nabla_x\phi)^2  
\nonumber \\
&+&\frac{1}{2\rho_0} (\rho - y)^2 + {m\over{2}} \left((\nabla_t y)^2 + V_d^2(\nabla_x y)^2 \right)  
 \nonumber \\
&-&  u_P \cos\left(2\pi y(x,t)\right) -F y(x,t)] \label{H},
\end{eqnarray}
 where all the variables are periodic in the imaginary time $t\geq 0$ with the period $\beta=1/T$ (units $\hbar=1$, $K_B=1$); the core density $ \rho$ and the superfluid phase $\phi$ are canonically conjugate variables, with
$\rho'=\rho -y$ being the local superfluid density;
the derivatives $\nabla_{t,x} y$, $\nabla_{t,x} \phi$ are understood as finite differences in the discretized space-time lattice (with 200 time slices and
$x=0,1,2,...,L$ in units of $b$), with 
$\nabla_{t,x} \phi$ defined modulo 2$\pi$ (in order to take into account phase-slips); $n_0, \rho_0$ stand for the average filling factor ( we choose $n_0=1$) and the bare superfluid stiffness, respectively, with the bare speed of first sound taken as unity.

The first two terms in Eq.(\ref{H}) describe the superfluid response of the core \cite{sclimb}, and  the third term accounts for the superclimb effect \cite{sclimb} -- building  the dislocation edge so that the core climb $y\to y \pm 1, \pm 2, ...$ becomes possible by delivering matter $\rho \to \rho \pm 1,\pm 2, ...$, respectively, along the core \cite{sclimb}. The dislocation is assumed to be attached to large superfluid reservoirs at both ends, with spatially periodic boundary conditions for the supercurrent.

The terms $\propto m $ in Eq.(\ref{H}) account for the elastic response of the string, with $m$ and $V_d$ standing for the  effective mass of the dislocation core (per $b$) and the bare speed of sound, respectively. Since the main source of kinetic energy are supercurrents, we have left out the
term $\sim (\nabla_t y)^2$ in Eq.(\ref{H}). 
The parameter $m$ in Eq.(\ref{H}) is not actually a constant.  It contains a contribution from the Coulomb-type interaction potential $ \propto 1/|x|$ between jogs (or kinks, cf. \cite{EPL}) separated by a distance $x$ \cite{Hirth}. Accordingly, $m$ has a logarithmic divergent factor 
  with respect to a wave vector $q$ along the core $m(q) =  m_0\cdot\left[1+ U_C \ln\left(1+ \frac{1}{(bq)^2}\right)\right]$, where $m_0$ is of the order of the atomic \he4 mass and $U_C \sim 1$ is a parameter characterizing the strength of the interaction \cite{EPL,JLTP}. 
  In solid \he4, the zero-point fluctuation parameter $K=\pi \hbar/(4m_0 bV_d) \sim 1$  \cite{EPL}, which justifies the necessity of implementing full quantum-thermal analysis. We present our numerical results for $U_C=1, V^2_d=5, K=1$. 
It is important to note that the main results are not qualitatively  sensitive to the long-range interaction. 

As discussed in Ref.\cite{us_PRL_2011}, in the case of a single dislocation in an ideal crystal the linear force density $F\approx b\sigma$ (ignoring spatial indices) is determined by applied
chemical potential in the setup \cite{Hallock_2010}. 
Here we consider $F$  in Eq.(\ref{H}) determined by 
  external stresses $\sigma \sim 0.1-0.01$bar usually existing in solid \he4 ( $\sigma=F$ in units of $b$). Such stresses are inhomogeneous. We argue that they
are responsible for the roughening effect \cite{us_PRL_2011} in real samples, and their inhomogeneity may actually wash out the effect of the applied
chemical potential.

\begin{figure}[h]
\begin{minipage}{20pc}
\includegraphics[width=20pc]{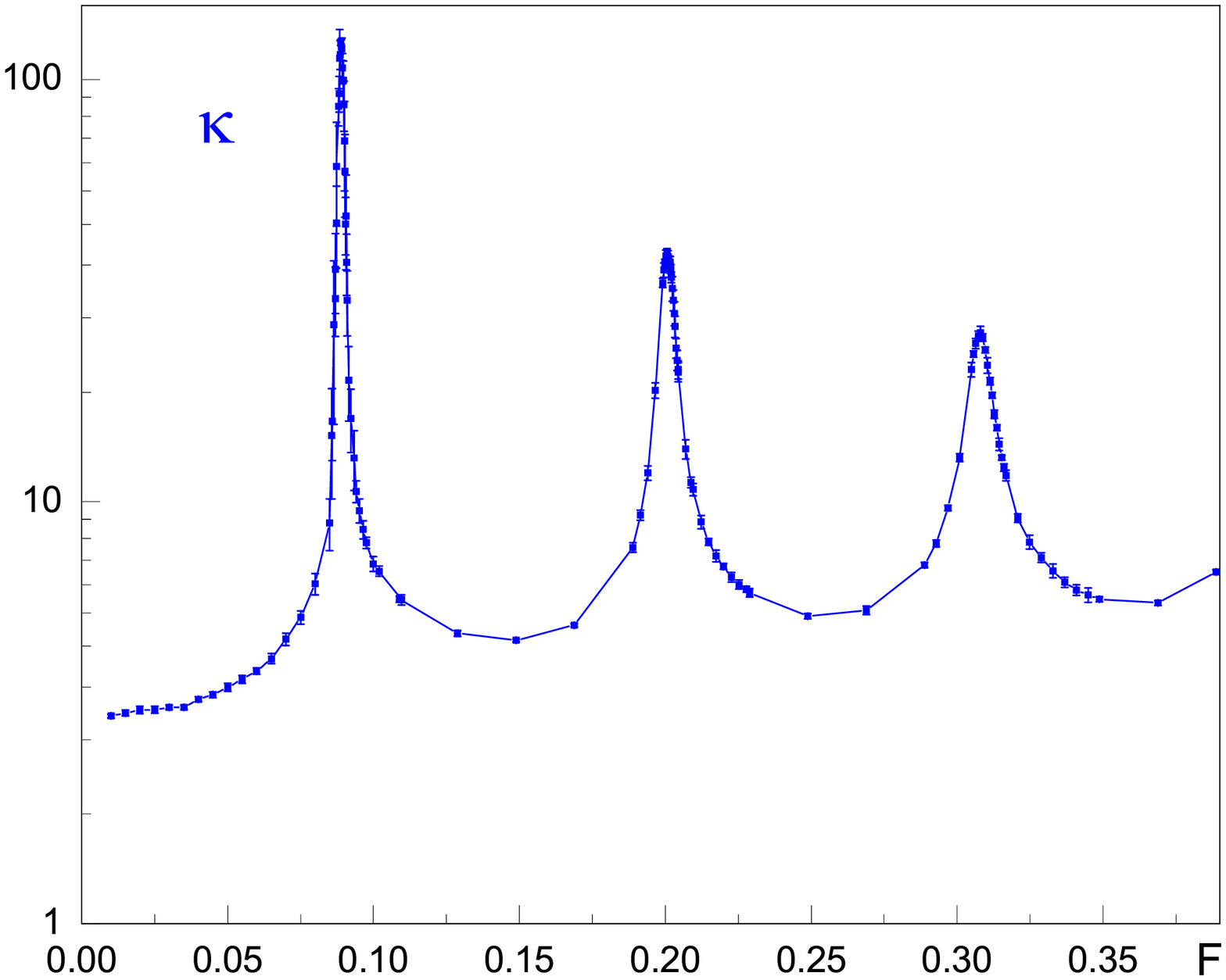}
\vspace{-0.5cm}
\caption{\label{kappa}(Color online)  Superfluid compressibility of the dislocation core. The resonant-type peaks occur at the thresholds $F=F_c(L,n),\, n=1,2,3$ for $n$ jog-antijog pairs creation.The parameters are $L=56, \, T=0.05, u_P=3.0$.}
\end{minipage}\hspace{2pc}%
\begin{minipage}{20pc}
\includegraphics[width=20pc]{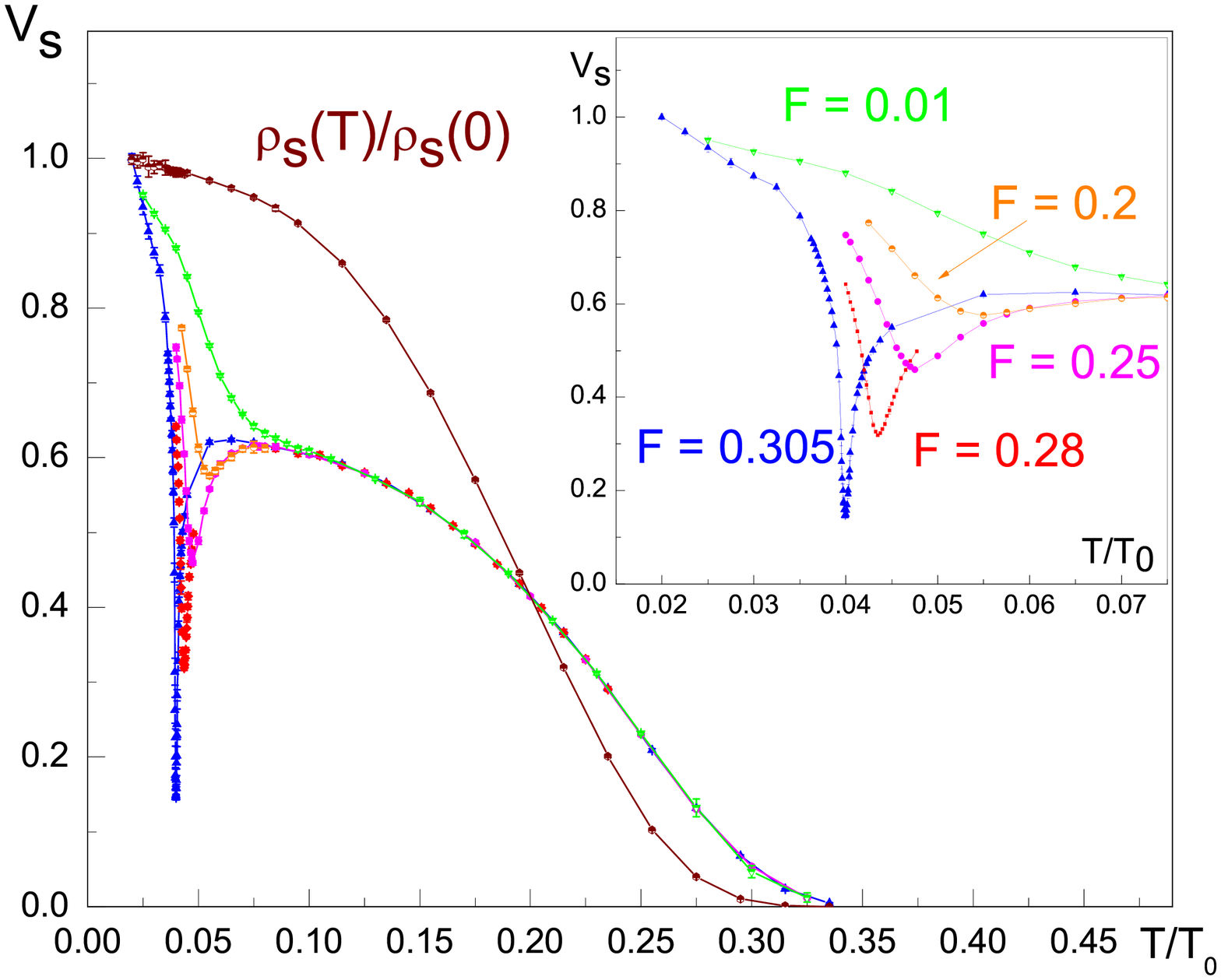}
\vspace{-0.5cm}
\caption{\label{VsT}Renormalized superfluid stiffness $\rho_s(T)$ and the velocity $V_s(T,F)$ of first sound  normalized by their respective low-$T$ values for different 
$F $ (shown on the inset), $L=30$, $u_P=3.0$. Inset: the region of the dip (cf. Fig.4 of Ref.\cite{Hallock_2010}) showing its shifting with $F$.}
\end{minipage} 
\end{figure}

Monte Carlo simulations  have been conducted with the Worm Algorithm (WA) \cite{WA} for the superfluid part of the action, with the Peierls term treated within the Villain approximation similarly to Refs. \cite{EPL,JLTP}.
The  renormalized superfluid stiffness $\rho_s(T,F)$ and compressibility $\kappa(T,F)$ have been calculated in terms of the windings of the dual variables \cite{WA,EPL,JLTP}. 
 No significant effect of the bias $F$ was found on $\rho_s(T,F)$, and thus we consider $\rho_s(T,F)=\rho_s(T,0)\equiv \rho_s(T)$. In contrast, $ \kappa(T,F)$ does experience quite dramatic renormalization as found in Ref.\cite{us_PRL_2011} and demonstrated in Fig.~\ref{kappa}.Thus,the renormalized speed of first sound
\begin{equation}
V_s(T,F)=\sqrt{\frac{\rho_s(T)}{\kappa(T,F)}}
\label{Vsf} 
\end{equation}
exhibits the dip \cite{us_PRL_2011}. 
According to the Landau criterion such speed limits the critical superflow rate along the dislocation core.

The bare stiffness $\rho_0(T)$ vanishes above some temperature $T_0$ comparable to the bulk $\lambda$-temperature ($\sim 1-2$K). We have used $T_0=0.2$ (in the dimensionless units as a fraction of the Debye temperature $T_D$ for the first sound), and considered low temperatures --  such that $\rho_s(T)$ stayed unchanged within 1-10\% of its $T=0$ value.  In other words, the thermal length $L_T \approx \rho_s(0)/T $, above which $\rho_s(T)$ becomes suppressed, is the largest scale in the problem. 
\begin{figure}
\centerline{\includegraphics[angle =0,width=0.9\columnwidth]{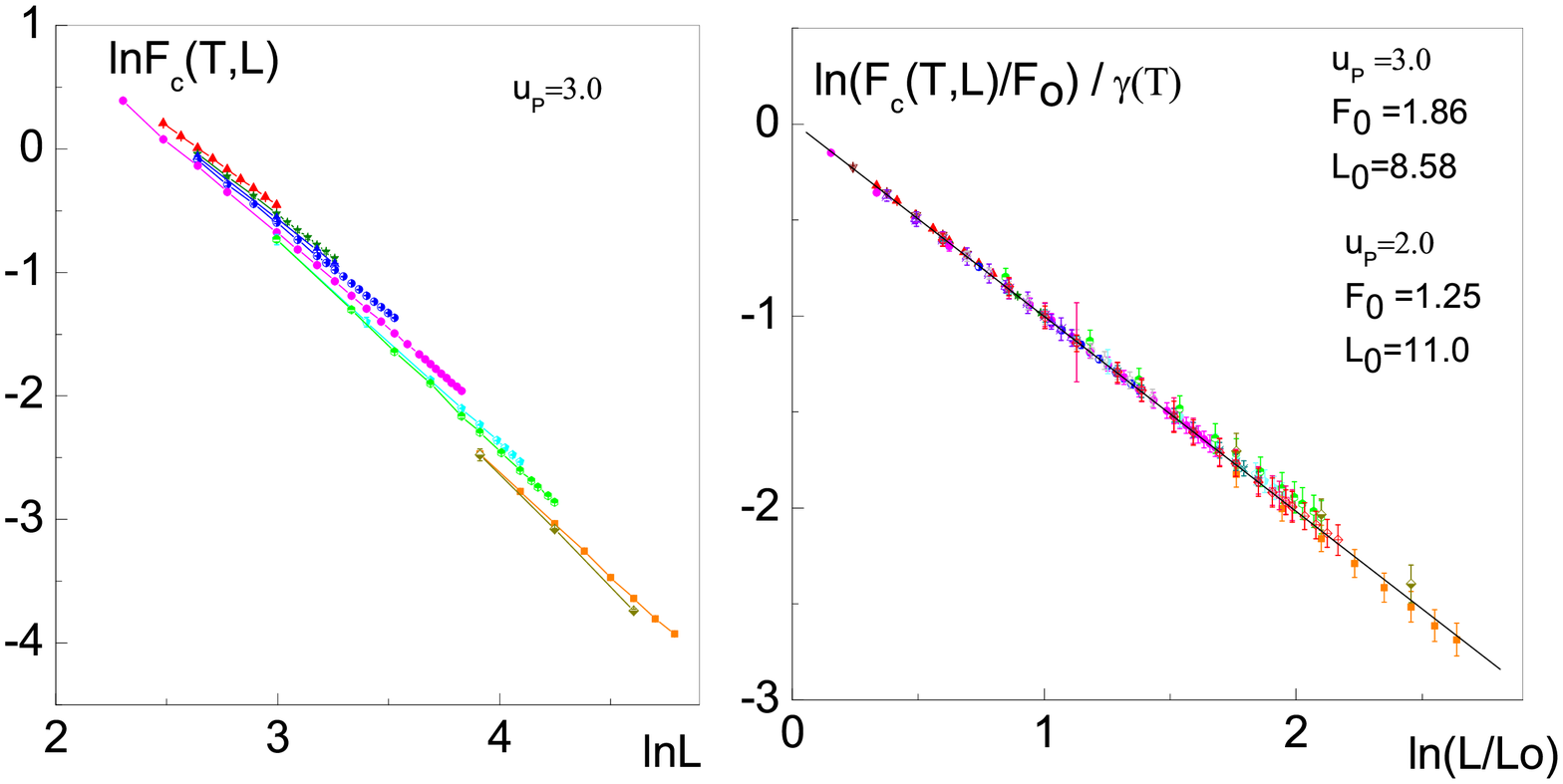}}
\vspace{-2.5cm}
\caption{(Color online) The left panel: power law dependence of the critical stress for several temperatures $T=0.006 -0.013$ (in the dimensionless units).
The right panel: the same dependencies for all temperatures collapsed to a single master curve. The parameters $F_0,L_0$ are shown for two different
values of the Peierls potential strength. Error bars are shown for all points. Their bigger values on the right panel are due errors in values of $\gamma$.}\label{Fcol}
\end{figure}
Jog-antijog pairs as quantum objects can be created spontaneously by a {\it macroscopically} small stress $F \geq F_c\propto 1/L$ applied to a superclimbing dislocation of length $L$ -- analogous to the creation of kink-antikink pairs along a stressed gliding dislocation \cite{Hirth,Petukhov}. In Ref.\cite{us_PRL_2011} we have found that such an instability leads to a first-order phase transition even at finite temperature $T$ between two phases of the dislocation -- {\it smooth} and {\it rough}.
Such transition is characterized by hysteretic behavior once dislocation length exceeds a certain threshold $L_h$ growing with $T$ \cite{us_PRL_2011}. At sizes $L<L_h$
the roughening is characterized the resonant-type increase of the dislocation compressibility $\kappa$, Fig.~\ref{kappa}. As a result, speed of the first sound along the dislocation core
exhibits a strong suppression shown on Fig.~\ref{VsT} for the first resonant peak. As can be seen, the strongest peak in Fig.~\ref{kappa} corresponds to the single pair creation. Thus, in what follows we will be ignoring the higher order processes.

\subsection{The critical stress for the dip}
The resonant-type increase of the compressibility occurs at some value of the critical stress $F=F_c$ which depends 
strongly on dislocation length and less strongly on temperature. With good accuracy such dependence
has been found to be
\beq
F_c(L,T)=F_0 \left(\frac{L_0}{L}\right)^{\gamma(T)}
\label{Fc}
\eeq
where $F_0,\, L_0$ are constants independent of temperature, and $\gamma(T)$ varies from unity in the limit of $T=0$ to $\gamma \approx 1.8$ 
at temperatures approaching the temperature comparable to the energy of a jog-antijog pair $\Delta$. Fig.~\ref{Fcol} demonstrates such power law
dependence for several temperatures. The constants $F_0,\, L_0$ depend on the strength of the Peierls potential $u_P$, that is, on the parameters
of the jog-antijog pair. In terms of the dimensional values, $F_0$ is determined by a typical Peierls stress $\sigma_P \sim 10-100$bar ( $F_0 \sim b \sigma_P$)
and $L_0$ is given by a typical jog-size. In what follows we will be considering stresses $F$ which are much smaller than $F_0$.

\subsection{The dip temperature}
In the case when the free segment length is determined by the mean distance between the impurities as presented in Eq.(\ref{L}), the condition
of the resonance (\ref{Fc}) can be eventually met for any (small) value of the external stress $F$. As will be shown below
the temperature $T_d$ at which this occurs depends weakly (logarithmically) on the actual value of $F$. Indeed, substituting Eq.(\ref{L}) into
Eq.(\ref{Fc}) and solving for $T$ in the approximation $\gamma\approx const$ we find
\beq
T_d= - \frac{E_a}{\ln\left[L_0(F_0/F)^{1/\gamma}X_3 -x_d\right]}.
\label{Tc}
\eeq
This equation is valid when $0< L_0(F_0/F)^{1/\gamma}X_3 -x_d <1$. Given $x_d <<1$, this translates into the wide
range of the stresses
\beq
(X_3 L_0)^\gamma < F/F_0 < (X_3 L_0/x_d)^\gamma.
\label{ran}
\eeq
Since $X_3/x_d \sim 1, \,\, L_0 \geq 1$ and $F<<F_0$, it is reasonable to ignore $x_d$ in the denominator of Eq.(\ref{Tc}). Then, the dip temperature
becomes much less than the activation energy
\beq
T_d= - \frac{E_a}{\ln\left[L_0(F_0/F)^{1/\gamma}X_3\right]}  ,
\label{Tc2}
\eeq
provided $\ln\left[L_0(F_0/F)^{1/\gamma}X_3\right] << -1$. 
It is important that $T_d$ depends only logarithmically on \hee3 concentration and the values of
the random stresses. As an example, taking $X_3=10^{-7}$ and using $F/F_0 = 10^{-4}$, $L_0=10,\, \gamma \approx 1.5$, $E_a=0.5$K, we find
$T_d \approx 0.07$K, with the log-factor being about 8. \hee3 concentration could be varied in a wide
range of values: from $10^{-12}-10^{-15}$ in, practically, \hee3 free samples to $10^{-4}-10^{-5}$. Thus, despite the logarithmic  dependence, a quite significant
shift in the position of the dip can be observed. We find important studying such dependence experimentally. For example, for $X_3 \sim 10^{-12}$ and presuming the same
protocol for preparing the sample, so that, the internal stresses remain essentially the same, one finds the dip
temperature reduced by more than a factor of two.    

As discussed in Ref.\cite{us_PRL_2011}, the dip width is exponentially decreasing with the dislocation length (and eventually the resonant
behavior turns into the hysteretic one). This conclusion is valid for a single dislocation. In the case of the dislocation forest with 
spatial fluctuations of the frozen in stresses the dip depth should rather be determined by  a sort of inhomogeneous broadening -- controlled
by variations of $F$ over a sample. If the average (mean square) fluctuation of the stress $\delta F $ is smaller than the mean square stress
$\langle F \rangle$ itself, the estimate of the depth of the inhomogeneously broadened dip becomes
\beq
\delta T_d= \frac{T_d}{\gamma |\ln\left[L_0(F_0/\langle F \rangle)^{1/\gamma}X_3\right]|} \frac{\delta F}{\langle F \rangle} .
\label{dTc2}
\eeq
 Given that for practical values of the parameters the logarithm is of the order of 10, we conclude that $\delta T_d << T_d$ even if $\delta F \sim \langle F \rangle$.

\section{Discussion}
Here we have proposed that \hee3 impurities in combination with random frozen in stresses are controlling the dip-anomaly in the superflow rate through solid \he4 observed in Ref.\cite{Hallock_2010}. 
For low density of dislocations the length of the dislocation segments $L$ undergoing the stress induced roughening \cite{us_PRL_2011} is controlled by
 the inter-impurity distance along the dislocation cores rather than by the cross-pinning of dislocations.
Due to the frozen in spatial fluctuations of stresses, there should generically always be segments meeting the condition
for the resonant-type jog-antijog pair creation leading to a strong suppression of the flow. The exponential dependence of $L$ on temperature insures
that such resonances occur in a narrow temperature range (see Eq.(\ref{dTc2})) around the logarithmically suppressed temperature (\ref{Tc2}). Thus, in contrast to a single
dislocation in ideal crystal,
in real samples the dependence on the average stress becomes much weaker. This may be the reason for a very weak dependence of the flux rate in the dip region
found in Ref.\cite{Hallock_2011}, Fig.20. Under these circumstances we find important studying experimentally the dependence of the dip temperature on \hee3 concentration.

\section{Acknowledgments}
We thank Robert Hallock for discussing details of his experiment.
This work was supported by  the National Science Foundation, grant No.PHY1005527,  PSC CUNY, grant No. 63071-0041, by the CUNY HPCC under NSF Grants CNS-0855217 and CNS - 0958379.

\end{document}